\documentclass[review]{elsarticle}
\usepackage{graphicx}

\journal{Journal of \LaTeX\ Templates}

\begin{document}

\begin{frontmatter}

\title{Strain effects on optical properties of tetrapod-shaped CdTe/CdS core-shell nanocrystals}

\author[mymainaddress]{Yuanzhao Yao\corref{mycorrespondingauthor}}
\cortext[mycorrespondingauthor]{Corresponding author}
\ead{yao.yuanzhao@nims.go.jp}
\author[mymainaddress]{Takashi Kuroda}

\author[mysecondaryaddress]{Dmitry N. Dirin}
\author[mysecondaryaddress]{Maria S. Sokolikova}
\author[mysecondaryaddress]{Roman B. Vasiliev}

\address[mymainaddress]{Photonic Materials Unit, National Institute for Materials Science, 1-1 Namiki, Tsukuba, Ibaraki, 305-0044, Japan}
\address[mysecondaryaddress]{Faculty of Materials Science, M. V. Lomonosov Moscow State University, 119991 Moscow, Russia}

\begin{abstract}
The exciton states of strained CdTe/CdS core-shell tetrapod-shaped nanocrystals were theoretically investigated by the numerical diagonalization of a configuration interaction Hamiltonian based on the single-band effective mass approximation. We found that the inclusion of strain promotes the type-II nature by confining the electrons and holes in nonadjacent regions. This carrier separation is more efficient than that with type-II spherical nanocrystals. The strain leads to a small blue shift of the lowest exciton energy. Its magnitude is smaller than the red shift induced by increasing the shell thickness. Moreover, the larger arm diameter reduces the influence of both shell thickness and strain on the exciton energy. Considering the broken symmetry, a randomness induced carrier separation was revealed which is unique to a branched core-shell structure. The calculated shell thickness dependence of the emission energy agrees well with available experimental data.
\end{abstract}


\end{frontmatter}


\section{Introduction}

Tetrapod-shaped nanocrystals (NCs) made of II-VI semiconductors have attracted a great deal of interest since the first report on their solution-processed chemical synthesis in 2000 \cite{manna2k}. Studies have been reported on tetrapods made of CdSe \cite{manna2k, fiore09}, CdS \cite{fiore09}, CdTe \cite{fiore09, manna03, giorgi05, tari05, malkmus06, tari06, morello08, goodman10, vasiliev09}, ZnTe \cite{fiore09}, and their core-shell combinations (core-shell tetrapods denoted hereafter as csTPs) \cite{peng05,talapin07,choi09,vitukhnovsky09,vasiliev09a,dirin11}. In addition to these synthesis and characterization studies, there have been reports on the application of tetrapods to photovoltaic cells \cite{li09,sun03,Zhou06} and single electron transistors \cite{cui05}.
The tetrapod-shaped NCs also attracted the interest of the theoretical investigations, which include calculation of the one-particle states of bare tetrapods by the semiempirical pseudopotential method \cite{li03,milliron04}  and effective mass approximation \cite{tari05,Morello11calc} as well as the calculation of exciton states by Hartree approximation \cite{lutich10,Mauser10,Mauser08}. We recently calculated the exciton states of II-VI bare tetrapods using a configuration interaction approach and obtained a good agreement with reported experimental data \cite{sakoda11, Yao2013bareTP}. But to the best of our knowledge, there has been no theoretical investigation of the exciton states of csTPs.

Comparing with the bare tetrapod structure, the study of csTPs is more complicated. On the other hand, a wide variety of interesting results can be expected as regards the electronic and optical properties of csTPs. First, in accordance with the band alignments of their constituent materials, synthesized csTPs may have a type-I or type-II band structure, that is, the electrons and holes tend to be localized in the same or different positions in the csTPs, respectively \cite{peng05,vasiliev09,dirin11}. Specifically, the type-II structure can have a smaller effective band gap than that of either of its constituent semiconductors. Therefore, with the type-II csTP, there is great potential for optimizing their absorption and luminescence properties for various applications by tuning their geometric parameters.

Second, it is well-known that structural symmetry plays an important role regarding the nature of carrier wave functions and also crucially influences the optical properties of the NCs. The difference between the optical properties induced by the tetrahedral symmetry of csTPs and those of the well-studied type-II core-shell NCs with spherical symmetry is interesting for fundamental investigation.

Third, the large lattice mismatch between the constituent materials, e.g. CdTe and CdS, leads to non-negligible strain in the csTP. The strain strongly affects the band structure of the csTPs and so is important to the investigation of the optical properties of csTPs.

To clarify the above three points, in the present study, we applied the same theoretical method to the exciton states of CdTe/CdS csTP NCs with different dimensions. A comparison with type-II spherical NCs revealed a uniquely efficient carrier separation in the csTPs. In addition to studying excitons in csTPs with perfect tetrahedral symmetry, we also discussed the effect of broken symmetry. For the CdTe/CdS csTPs with a type-II band structure, the absorption spectra are mainly contributed by the high-energy exciton states, which are beyond the range of the present investigation.


\section{Theory}

\begin{figure}[htbp]
\centering\includegraphics[width=14cm]{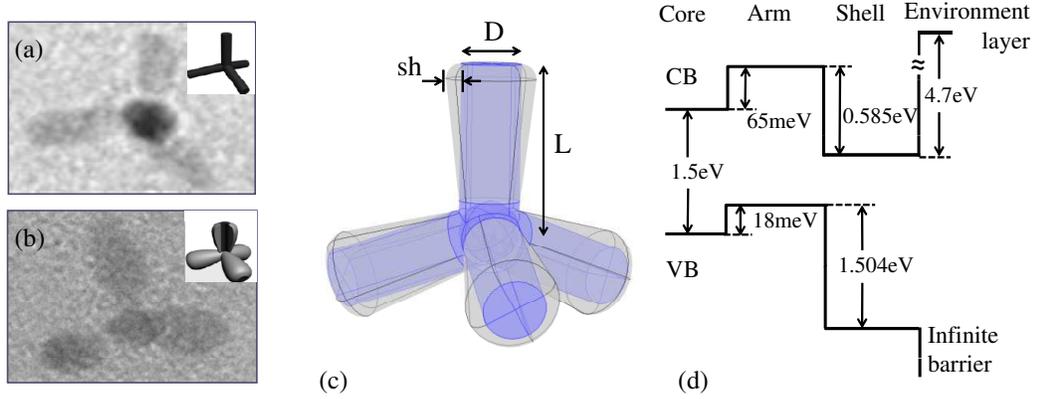}
\caption{SEM images of (a) bare tetrapod and (b) core-shell tetrapod. (c) Structure of the CdTe/CdS core-shell tetrapod assumed in the calculation. It consists of a spherical central core (CdTe with zinc blende structure), four cylindrical arms (CdTe with wurtzite structure), and four CdS shells covering the arms. The maximum shell thickness, and the diameter and length of the arms are denoted by $sh$, $D$ and $L$, respectively. The diameter of the central core is assumed to be the same as $D$.
(d) Energy band diagrams for electrons and holes. The confinement potential height of CB is assumed to be the same as the electron affinity. 4.7 eV is assumed for the CdS electron affinity, which is the sum of the electron affinity (4.18 eV) for CdTe with the zinc blende structure \cite{Bechstedt88} and the CB offset between CdTe and CdS \cite{dirin11}, whereas 1.5 eV is assumed for the zinc blende CdTe bandgap \cite{tari05}. An infinite potential barrier is assumed for the VB. As for the band offset between the CdTe with the zinc blende and wurtzite structures, we used 65 meV for the CB and 18 meV for the VB, which were obtained by theoretical calculation \cite{wei2k}. For the effective masses of the electrons ($m^{*}_{e}$) and heavy holes ($m^{*}_{h}$), we assumed $m^{*}_{e}=0.11\times m_{0}$ and $m^{*}_{h}=0.69\times m_{0}$ for CdTe \cite{long97} and $m^{*}_{e}=0.18\times m_{0}$ and $m^{*}_{h}=0.7\times m_{0}$ for CdS \cite{Mauser08}, where $m_{0}$ is the genuine electron mass.}
\label{fig1}
\end{figure}

Fig.~\ref{fig1}(a) shows the three-dimensional structure of the CdTe/CdS csTP that we assumed in our numerical study in accordance with the observation reported in Ref.\ \cite{vasiliev09a}. It consists of a spherical central core (CdTe with zinc-blende structure), four cylindrical arms (CdTe with wurtzite structure), and four CdS shells (CdS with wrutzite structure) covering the lateral surfaces of the arms. The shells on the csTP arms are isolated from each other. The maximum shell thickness, and the diameter and length of the arms are denoted by $sh$, $D$ and $L$, respectively.

We firstly assumed that the shells on all the arms were the same except for their spatial orientation, thus this csTP model exhibited perfect tetrahedral symmetry as with the bare tetrapods in Ref. \ \cite{sakoda11}. The band diagrams for the electrons and holes of the bulk CdTe and CdS are shown in Fig.~\ref{fig1}(b). When the effect of strain is taken into consideration, these band diagrams will be modified due to the distortion of the lattice framework.

To calculate the strain distribution over the entire csTP heterostructure, we used the approach described in Ref.\ \cite{Povolotskyi06} for a freestanding structure. We applied the continuum elasticity approximation, which assumes that the heterostructure has a coherent interface where the lattice points of the constituent materials match each other perfectly.

To fulfill the interface matching requirement, we initially assumed an interface matching configuration with unstrained CdTe arms and a deformed shell.
Considering the gradual growth of the thin shell on the arm, we assumed the CdTe arm to be the unstrained substrate \cite{Povolotskyi06}.
The central core was assumed to be a part of the unstrained substrate due to the lattice similarity between the wurtzite CdTe arm and zinc-blende CdTe core at their interface.
The deformation is denoted by $\bf u^0$ ($\bf u^0= 0$ in the arm), which produces anisotropic initial strain $\varepsilon^{0}_{xx}=\varepsilon^{0}_{yy}=(a^{A}-a^{S})/a^{S}$, and $\varepsilon^{0}_{zz}=(c^{A}-c^{S})/c^{S}$,
in which the superscript A(S) denotes the arm (shell)domain in the csTP model,
and $a$ and $c$ are the lattice constants of wurtzite crystals. The lattice constants used in our calculation were $a=0.457$ nm, $c=0.747$ nm \cite{Stuckes64} and $a=0.4136$ nm, $c=0.6713$ nm \cite{LB79}
\ for wurtzite CdTe and CdS, respectively.
With the interface matching assumption, no rotation was considered for the initial strain. Thus the initial strain tensor had zero off-diagonal elements. The initial strain can be expressed as:
\begin{equation}
\varepsilon^{0}_{ij}=\frac{1}{2}
( \frac{\partial u^{0}_{i}}{\partial x_{j}}
+ \frac{\partial u^{0}_{j}}{\partial x_{i}} )   \quad i,j = x,y,z
\end{equation}
where $u_{i}$ is a component of the deformation $\bf u^{0}$.

It should be noted that the interface matching configuration is not an equilibrium configuration.
The system will deform with respect to the interface matching configuration and relax to equilibrium with the lowest elastic energy, while maintaining the interface matching. During this relaxation, the deformation vector is defined as ${\bf u}$. The consequent elastic strain is  related to the components of ${\bf u}$:
\begin{equation}
\varepsilon^{e}_{ij}=\frac{1}{2}
( \frac{\partial u_{i}}{\partial x_{j}}
+ \frac{\partial u_{j}}{\partial x_{i}} )   \quad i,j = x,y,z
\end{equation}

Because the total strain distribution that we needed is generated by the total deformation ($\bf u^{0}+u$) from the initial unstrained configuration to the equilibrium configuration, the total strain of the system is:
\begin{equation}
\varepsilon_{ij}=\varepsilon^{e}_{ij}+\varepsilon^{0}_{ij}\delta_{ij} \quad i,j = x,y,z
\end{equation}
where $\varepsilon^{0}_{ij}=0$ in the arm domain, and $\delta_{ij}=1$ (or $\delta_{ij}=0$) if $i=j$ ($i\neq j$).
The elastic energy is:
\begin{equation}
W=\int \frac{1}{2} \Sigma_{ijkl}C_{ijkl}
\varepsilon_{ij} \varepsilon_{kl}dV,\quad  i,j,k,l = x,y,z
\end{equation}
where $C_{ijkl}$ is the anisotropic elastic modulus tensor for the wurtzite structure, taken from the Landolt-B\"{o}rnstein database \cite{LB79}.

The elastic energy minimization was implemented with the finite element method software COMSOL Multiphysics based on the virtual work principle \cite{Kuo08}. The deformation and consequent strain distribution can be obtained for the equilibrium configuration.
To enhance the computational efficiency and accuracy, we calculated the strain distribution in the "core + one arm" region of the csTP according to its symmetry.
Because interface matching was maintained during the calculation, the continuous deformation of the constituent materials was employed as a boundary condition at the interface ($\bf {u}^{A}\mid_{interface}=\bf u^{S}\mid_{interface}$).
 For the uniqueness of the solution, a fixed core (${\bf u}=0$) was assumed as a constraint to prevent the translation or rotation of the structure.
 The other outer boundaries were specified as free surfaces due to the zero external force assumption for freestanding csTPs.

The strain induced modification of the lowest conduction band (CB) $V_{e\varepsilon}$ and the highest valence band (VB) $V_{h\varepsilon}$ can be evaluated using the strain-related Hamiltonian for a wurtzite semiconductor found in Ref. \ \cite{Adachi05}. Considering the single-band calculation in the present study, we found that the band modification of the electron in question and the heavy hole states were:

\begin{equation}
V_{e\varepsilon}({\bf r}_{e})=
a_{cz}\varepsilon_{zz}+
a_{ct}(\varepsilon_{xx}+\varepsilon_{yy})
\end{equation}
\begin{equation}
V_{h\varepsilon}({\bf r}_{h})=
(D_{1}+D_{3})\varepsilon_{zz}+
(D_{2}+D_{4})(\varepsilon_{xx}+\varepsilon_{yy})
\end{equation}
where $a_{cz}$ and $a_{ct}$, respectively, are the deformation potentials of CB along the c-axis and transverse to the c-axis of wurtzite materials, $D_{i}(i=1\sim4)$ are the deformation potentials of VB.
For the wurtzite CdTe arms, the deformation potentials were derived from those of zinc-blende CdTe \cite{VanDeWalle89} with the quasi-cubic approximation \cite{Adachi05}. The idea of this approximation is based on the similarity between the wurtzite structure along the [0001] direction and the zinc-blende structure along the [111] direction. For the wurtzite CdS shell, the deformation potentials were taken from Ref. \ \cite{LB79,park11}.

The strain induced band-edge shifts play a role in the potential modification of the heterostructure. This modification can be expressed as an extra potential term in the single-particle Schr\"{o}dinger equation:

\begin{equation}
 H_{i}\varphi_{i}(\textbf r_{i})\equiv\left\{-\frac{\hbar^2\triangle_{i}}{2m^*_{i}}+V_{i}(\textbf r_{i})+V_{i\varepsilon}(\textbf r_{i})\right\}\varphi_{i}(\textbf r_{i})=E_{i}\varphi_{i}(\textbf r_{i}), \quad i = e,h,
 \label{eq1}
\end{equation}
where
$\triangle$ is the Laplace operator,
$V_{i}$ is the band offset of unstrained CB and VB,
$\varphi_{i}$ is the envelope function of electrons and holes,
and $E$ is the energy eigenvalue.
$m^{*}$ is the isotropic effective mass assumed in our calculation.
The numerical calculations were performed with the finite element method using the commercial software COMSOL Multiphysics.

In the II-VI semiconductors, the heavy and light hole states are not degenerated, and the heavy hole states have lower kinetic energy due to their larger effective mass. Because we are only interested in the low-energy excitons, the VB state mixing is not concerned in the present paper. In addition, there was good agreement between the low-energy excitons of CdTe/CdSe core-shell spherical NCs calculated with single-band and multi-band theory \cite{Tyrrell11discardVBmix}. Thus the single-band approximation in the present paper is valid as long as we focus our discussion on low-energy excitons.

The obtained envelope functions and energy of the low-energy electron and hole states are utilized to form pair states for the calculation of excitons. The calculation of exciton states using configuration interaction approach followed the same procedures described in our previous study \cite{sakoda11}.
When the defect is not concerned, our present calculation method is suitable for the tetrapod-shaped core-shell nanocrystals with other materials in the strong confinement regime, which ensure the validity of configuration interaction approach and sufficient convergence in the numerical calculation.

\section{Results and Discussion}

\begin{figure}[htbp]
\centering\includegraphics[width=8cm]{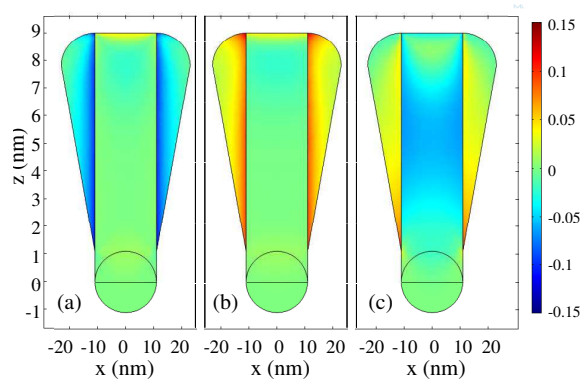}
\caption{Cross-section of the calculated strain components (a) $\varepsilon_{xx}$, (b) $\varepsilon_{yy}$, and (c) $\varepsilon_{zz}$ in one arm of a csTP with $sh=1.2$ nm.}
\label{fig2}
\end{figure}

Fig.~\ref{fig2} shows the distribution of the calculated strain components $\varepsilon_{xx}$, $\varepsilon_{yy}$, and $\varepsilon_{zz}$ in an x-z cross-section of one branch of a csTP with $sh=1.2$ nm. The CdTe arms are under compressive strain in all three directions due to the larger lattice constant.
On the other hand, the CdS shells are under tensile strain $\varepsilon_{yy}$ and $\varepsilon_{zz}$, but the $\varepsilon_{xx}$ in the x-z cross-section is compressive due to Poission's effect as shown in Fig.~\ref{fig2}(a).
The main features of the strain components in Fig.~\ref{fig2} agree with those of the InAs/InP core-shell nanowire \cite{boxberg10}, in which the core material has a larger lattice constant, as with the CdTe/CdS system in our calculation.

\begin{figure}[htbp]
\centering\includegraphics[width=10.5cm]{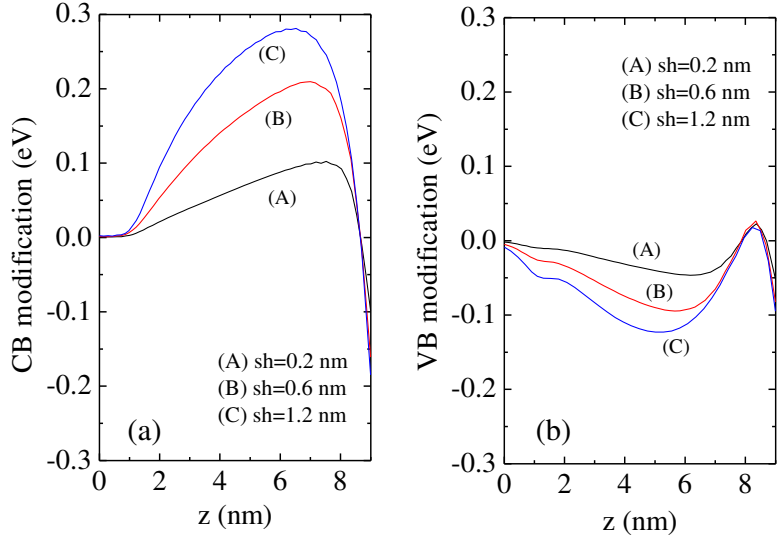}
\caption{
The strain induced band edge modification of (a) the CB and (b) the VB along the axial direction of the arms for csTPs ($D=2.2$ nm) with different shell thicknesses.
}
\label{fig3}
\end{figure}

The compressive strain in the arms is dominated by its $\varepsilon_{zz}$ components. Because the z direction component of the deformation potential of CdTe is negative for CB and positive for VB, the corresponding band edge energy shifts mainly have positive and negative values as shown in Fig.~\ref{fig3}.
This modification of VB is smaller than that of CB due to the smaller deformation potential in VB.
With increasing $sh$, the strain decreases in the shell and increases in the arm, leading to larger band modification in the arm. As a result, the effective band gap of the arm is larger and the type-II nature of the heterostructure band structure is more pronounced.

\begin{figure}[htbp]
\centering\includegraphics[width=14cm]{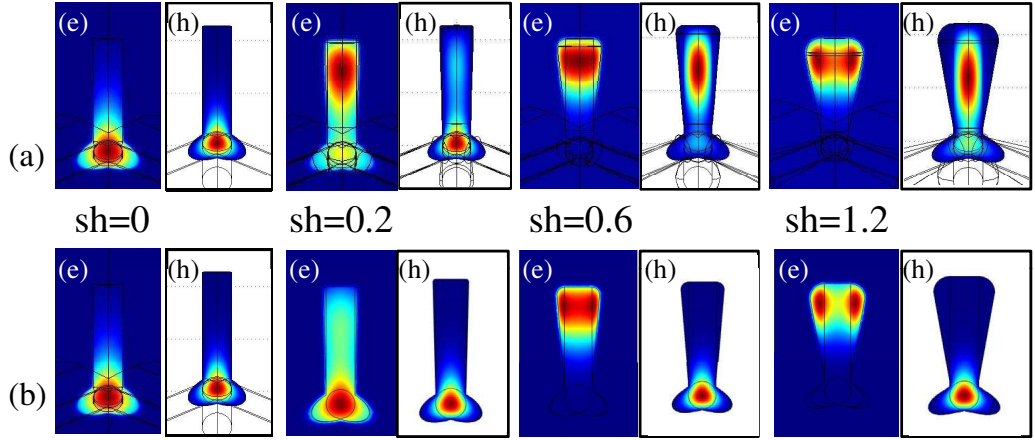}
\caption{Cross-section of the wave function of the lowest electron and hole states for csTPs with various shell thicknesses. Wave functions of csTPs (a) without and (b) with the strain effect are plotted for comparison. The labels (e) and (h) denote the wave function of electron and hole states, respectively}
\label{fig4}
\end{figure}

In a spherical type-II structure, we only need to be concerned with the carrier wave functions along the radial direction. When the shell thickness is infinite, the electrons and holes can be considered completely spatially separated. For real core-shell type-II spherical NCs, the finiteness of the shell thickness leads to a non-zero overlap of the confined electron and hole wave functions. Hence, a sufficiently large shell thickness is necessary for efficient carrier separation.

The situation in a csTP with tetrahedral symmetry is more complex than in a spherical heterostructure. Fig.~\ref{fig4} shows the effect of strain on the wave functions of the lowest electron and hole states, which mainly determine the nature of the lowest exciton state. The wave function of the lowest electron and hole states of csTPs have $A_{1}$ symmetry and are localized in the core region for a bare tetrapod ($sh=0$).
As $sh$ increases, the larger volume in the "CdTe arm + CdS shell" region attracts both the electrons and holes. The electrons are eventually distributed in the CdS shells for a sufficiently large $sh$, because of the smaller potential energy there. We notice that as the shell thickness increases to 1.2 nm, the electrons in the shell and the holes in the arm are still not completely separated.

The inclusion of strain leads to a more pronounced type-II band alignment in the CdTe/CdS csTPs. This phenomenon is consistent with the results for the core-shell type-II spherical NCs in which the core material has a larger lattice constant than the shell \cite{Cai2012csSphere, Fairclough2012csSphere,smith09}.
But we notice that the type-II nature of the strained csTPs induces the carrier separation more effectively. As shown in Fig.~\ref{fig4}(b), the strain induced band modification prevents carrier delocalization out of the core region. As $sh$ increases, electrons with larger kinetic energy can be redistributed in the shell with smaller potential energy, but the holes remain in the core region.
The electrons and holes can be considered completely separated according to their localization in the nonadjacent regions.

\begin{figure}[htbp]
\centering\includegraphics[width=8cm]{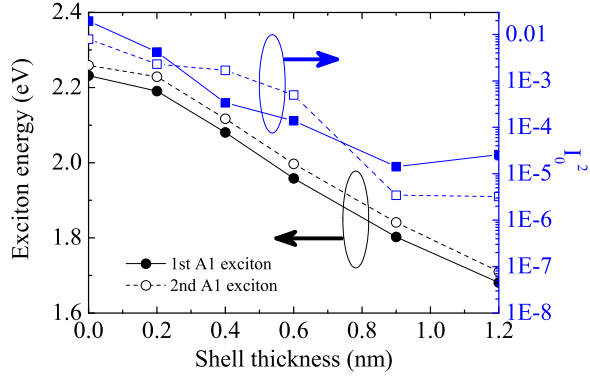}
\caption{Variation in the energy (circles) and the square of the overlap integral (squares) of the lowest (solid line) and the second lowest (dashed line) $A_{1}$ exciton as a function of shell thickness.}
\label{fig5}
\end{figure}

With the results of single particle states, we can discuss the exciton states in csTPs.
For all the $sh$ values in the present study, the lowest spin-singlet exciton has $A_{1}$ symmetry (optical active), which mainly consists of the lowest electron-hole pair state.
As $sh$ increased, a red shift in the exciton energy was observed for the type-II structure as a result of the decreasing confinement. Moreover, the oscillator strength of the lowest and the second lowest $A_{1}$ spin-singlet exciton quickly decreased due to the carrier separation as shown in Fig.~\ref{fig5}.
The oscillator strength of the lowest exciton decreased by $99\%$ when the shell thickness increased to 0.6 nm, reflecting the high efficiency of carrier separation in csTPs.
With a sufficiently large shell thickness, the luminescence of csTPs with perfect symmetry may be very weak. Therefore, we suppose that there are other contributions to the luminescence observed in the experiment, for example csTPs with broken symmetry, which we discuss below.


\begin{figure}[htbp]
\centering\includegraphics[width=10.5cm]{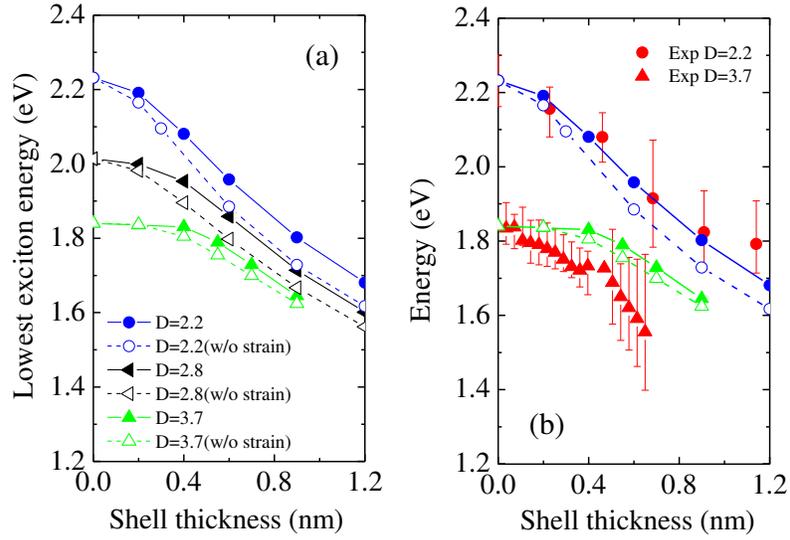}
\caption{(a) The shell thickness dependence of the lowest spin-singlet exciton energy of strained csTPs with different $D$ values. The corresponding data for unstrained csTPs are also plotted for comparison. (b) Comparison of the calculated lowest spin-singlet exciton energy with the experimental photoluminescence peak energy of strained csTPs. The data for $D=2.2$ nm and $D=3.7$ nm are denoted by circles and triangles respectively. The experimental results were shifted to align them with the calculated exciton energy at $sh=0$. The error bar indicates the full width at half maximum (FWHM) of the observed luminescence peak.}
\label{fig6}
\end{figure}

The effect of dimensions and strain on the energy of the lowest spin-singlet exciton is shown in Fig.~\ref{fig6} (a).
The inclusion of strain induces a small blue shift in the lowest exciton energy. This can be explained by the penetration of carrier wave functions into the CdTe arms with an enhanced band gap modified by the strain.
Meanwhile, the increasing $sh$ leads to a red shift with a magnitude larger than the influence of the strain. This is consistent with the calculation results for spherical core-shell type-II NCs \cite{Fairclough2012csSphere, park11}.
For csTPs with a large $D$, the energy of the lowest exciton is less tunable by $sh$, which is consistent with the results for CdTe/CdSe spherical dots reported in Ref. \ \cite{Tyrrell11discardVBmix}. On the other hand, the strain has less effect on the lowest exciton energy because there is less strain in an arm with a larger $D$.

Fig.~\ref{fig6} (b) compares calculation results and a previously reported experimental observation \cite{vasiliev09a}. To concentrate on the effect of shell thickness, the experimental results were shifted to align them with the calculation result at $sh=0$. When the large inhomogeneous broadening in the experimental results is taken into consideration, the $sh$ dependence of the calculated exciton energy agrees well with the photoluminescence data.


In an actual csTP specimen, broken symmetry may influence the carrier distribution and consequently the emission properties. Because it is impossible to study all kinds of randomness, here we analyzed the combination of parameters $D$ and $sh$, which have a dominant influence on the optical properties \cite{sakoda11}.
We studied the change induced by two kinds of modifications to csTP with $D=2.2$ nm  and $sh=0.9$ nm, and tried to identify the essential features of symmetry breaking in a qualitative manner:
For the first case, we modified one arm with a larger diameter $D'$ ($D' > D$) or a larger shell thickness  $sh'$ ($sh' > sh$) for the csTP; for the second case, we modified the same arm or two different arms of the csTP with a larger diameter $D'$ and a larger shell thickness $sh'$.

With a larger $D'$ or $sh'$ in only one arm, the low-energy electron and hole state tend to locate in the modified arm due to the increased confinement volume, which is consistent with the results described in Ref.\ \cite{Mauser08}.
We revealed that $D'$ and $sh'$  mainly influence the hole and electron distribution, respectively. On the basis of this result, we can design the hole and electron distribution by manipulating the parameters of each isolated branch of a csTP.

With the simultaneous modification of $D'$ and $sh'$ in two different arms, we found that the low-energy electron and hole states were localized on the corresponding different arms. This kind of randomness-induced carrier separation is unique for branched core-shell NCs, and cannot be found in core-shell spherical or rod systems.
With the simultaneous modification of $D'$ and $sh'$ on the same arm, the increased wave function overlap between low-energy electrons and holes is assumed to contribute to the luminescence in the experiment.

\section{Conclusion}

The exciton states of strained CdTe/CdS core-shell tetrapod-shaped NCs were investigated theoretically. The inclusion of the strain effect promotes the type-II nature of the band structure in CdTe/CdS csTPs. When compared with type-II spherical NCs, tetrahedral symmetry combined with the strain effect leads to more efficient carrier separation by confining the low-energy electrons and holes in nonadjacent regions. The strain effect induces a blue shift of the lowest spin-singlet exciton state. Increasing shell thickness leads to a red shift with a larger magnitude than the influence of strain, which is consistent with previous results for spherical type-II core-shell NCs. For csTPs with a larger $D$, both $sh$ and the strain have less influence on the energy of the lowest exciton. The shell thickness dependence of the calculated exciton energy agreed well with available experimental data. From a practical point of view, type II CdTe/CdS csTPs with charge separation are interesting for photovoltaic applications in devices with an active layer based on nanoparticles, wherein the charge separation occurs within the nanoparticle \cite{Rivest11}. So, the present calculation provides an opportunity to predict electronic properties and improve the effectiveness of charge separation.

The study of csTPs with broken symmetry revealed that electrons and holes can be confined in the same or different branches by manipulating the randomness. The randomness induced carrier separation into different branches of a csTP is unique for a branched core-shell heterostructure. When the electrons and holes localized in the same branch, we supposed the spatial direct transition in the branch contribute to the luminescence observed in the experiment. Tetrapods with broken symmetry provide evidence for the view that type II csTPs behave like four weakly connected quantum dots (each branch of the tetrapod) with the possibility of an electron remaining in a single dot for a long time. This might find interesting applications in nanoelectronics (e.g. as memory devices or elements for quantum computing).

\section*{Acknowledgement}
The author Y. Yao wishes to express his sincere appreciation to Prof. Kazuaki Sakoda in National Institute for Materials Science for stimulating discussion. Also Y. Yao wants to acknowledge Prof. Shun-Jen Cheng in National Chiao Tung University for the discussion about numerical calculation of strain.

%
%



%


\end{document}